\documentclass[twocolumn,aps,showpacs]{revtex4}
\usepackage{graphicx,psfrag}
\def\bc{\begin{center}}
\def\ec{\end{center}}

\def\beq{\begin{equation}}
\def\eeq{\end{equation}}

\begin{document}

%e-mail: ziegler@physik.uni-augsburg.de\\
%telephone: (49) 821 598 3244, FAX: (49) 821 598 3262

%\maketitle
%Title of paper
\title{Transport in finite graphene samples with random gap\\
%{\tiny fingraph.tex}
}

\author{K. Ziegler and A. Sinner}
\affiliation{Institut f\"ur Physik, Universit\"at Augsburg\\
D-86135 Augsburg, Germany}
\date{\today}

\begin{abstract}
We study the DC transport of finite graphene samples with random gap.
Using Dirac fermions to describe the low-energy physics near the Dirac point,
we employ a generalized Drude form for the conductivity. The latter is
constant for a vanishing average gap but always decreases with increasing
sample size for a nonzero average gap. The asymptotic conductivity of the
infinite sample is either nonzero if the average gap is smaller than a critical 
value or zero otherwise. Our results are in qualitative agreement with recent 
numerical calculations of Bardarson et al., Phys. Rev. B {\bf 81}, 121414(R) (2010).
\end{abstract}

\maketitle

%\section{Introduction}

Transport in graphene is a fascinating field in terms of experiments 
as well as theory \cite{novoselov05,zhang05,geim07,castroneto07b} and it provides an interesting
test ground for the validity of general theoretical concepts. A problem is that
experimental measurements and numerical simulations of transport properties
are performed on finite size samples. On the other hand, 
most analytic calculations are based on linear-response theory for infinite samples.
It is known from perturbative renormalization group calculations that disorder in graphene
can create its own finite length scale that describes a crossover from small scales
(weak disorder regime) to large scales (strong disorder regime) \cite{altland06,aleiner06}.
The perturbative renormalization 
group approach works only in the regime of weak disorder (i.e. on small scales), such that
the crossover behavior requires an alternative analysis. 
% beyond the perturbative renormalization group approach. 
An example is the random-gap model that is described by Dirac fermions with 
random mass. The renormalization of the average gap parameter ${\bar m}$ always scales to zero 
\cite{dotsenko82} , %,ludwig94,aleiner06,foster08}. 
indicating the total suppression of the gap by fluctuations.
Provided that the perturbative renormalization group approach is valid, its results imply that 
a random gap always scales to zero and graphene remains metallic for any strength of
gap fluctuations. This, of course, cannot be true
for a sufficiently large ${\bar m}$, where we expect an insulating behavior. The latter was
observed experimentally in case of hydrogenated graphene \cite{elias08}, in a recent numerical
simulation \cite{beenakker10} and analytically for
an infinite random-gap model \cite{ziegler09}. However, to compare the analytic theory with the experiment
and with numerical simulations \cite{beenakker10}, the finite-size crossover remains 
an open problem. In particular, the transition between metallic and insulating behavior
may be washed out by finite-size effects. An interesting result of the numerical 
calculation is that graphene is metallic with constant DC conductivity $\sigma_0=e^2/\pi h$
for vanishing average gap ${\bar m}$, regardless of the strength of the gap fluctuations, but 
it decreases with sample size for any nonzero average gap ${\bar m}$. This rises the question 
whether or not the conductivity vanishes for the infinite sample with any ${\bar m}>0$. 
This question will be addressed in the following.

We employ the linear response approach to study the transport properties
for the Dirac Hamiltonian $H$ with random gap term:
\[
H=i\nabla_1\sigma_1+i\nabla_2\sigma_2+m_r\sigma_3
\ .
\]
$\sigma_j$ is a Pauli matrix and the random gap $m_r$ has an uncorrelated Gaussian 
distribution with mean ${\bar m}$ and variance $g$.
Then the conductivity can be evaluated from the Kubo formula. The latter is applicable also to finite samples, 
provided that the boundary conditions are properly taken into account \cite{fujii07,nicolic01}. 
Then the conductivity is related to the average two-particle Green's function as in the 
infinite sample. 
%In this work the averaging is performed with respect to a uncorrelated Gaussian 
%distribution of a random gap with mean ${\bar m}$ and variance $g$. 
The average conductivity at the Dirac point becomes the simple scaling form \cite{ziegler09}
\beq
\sigma(\omega)=F({\bar m},\omega+2i\eta)\frac{e^2}{\pi h}
\ ,
\label{drude1}
\eeq
where $\omega$ is the frequency of the external field. 
It should be noticed that the conductivity depends on the fluctuations of the random gap 
only through the scattering rate $\eta$. 
This is a generalized Drude formula with the scaling function $F(z)$. The conventional
Drude formula has $F(z)\propto 1/z$. 
In case of graphene the DC limit $\omega=0$ of the % real part of the 
conductivity has the scaling form \cite{ziegler09} 
\beq
\sigma_0\sim % \frac{4\eta^2}{\pi(4\eta^2+{\bar m}^2)}\frac{e^2}{h}=
\left(1-\frac{{\bar m}^2}{4\eta^2+{\bar m}^2}\right)\frac{e^2}{\pi h}
%=\frac{a(m_c^2-{\bar m}^2)}{\pi m_c^2}\Theta(m_c^2-{\bar m}^2)\frac{e^2}{h}
\ .
\label{cond5}
\eeq
The scattering rate $\eta$ is real, and the conductivity vanishes for $\eta\to 0$.
A vanishing scattering rate can be understood here as the absence of quantum states,
which reflects the appearence of an effective gap.

%for ${\bar m}\le \sqrt{4\eta^2+{\bar m}^2}$. 
The scaling form of Eq. (\ref{cond5}) allows us to define a critical gap parameter as
$m_c=\sqrt{4\eta^2+{\bar m}^2}$ \cite{beenakker10} which itself depends on ${\bar m}$. 
Then the conductivity vanishes as one approaches ${\bar m}=m_c$.
In the infinite sample the scattering rate $\eta$ is determined by \cite{ziegler09}
\beq
\frac{1}{4\pi^2}\int_0^{2\pi}\int_0^{2\pi}\frac{1}{\eta^2+{\bar m}^2/4+k^2}dk_1dk_2
=\frac{1}{g}
%\approx\frac{1}{4\pi}\log\left[1+\lambda^2/(\eta^2+{\bar m}^2)\right]
\ ,
\label{spe1}
\eeq
where $g$ is the variance of the fluctuating gap. $\eta$ is automatically zero if there is 
no real solution $\eta$ for Eq. (\ref{spe1}) \cite{saddlepoint}. 

The scaling relation in Eq. (\ref{drude1}) and a nonzero DC conductivity are consequences 
of a scale-invariant behavior of the system due to a massless mode \cite{ziegler09,ziegler98}. 
The latter reflects a diffusive behavior of the quasiparticles, where the diffusion coefficient 
is finite and proportional to $1/\eta$ in the model under consideration. 
Diffusion is also possible in a finite
(confined) system with the restriction that the diffusion time is limited by the
sample size. However, the restriction can be avoided by applying periodic or
quasiperiodic boundary conditions, where the diffusing particles can leave the
sample on one side and re-enter it from another side. This corresponds with a physical
system in an electric circuit as it is described by the Kubo formula. 

Our previous calculations can be extended to finite samples with $N\times N$ sites.
The continuous wavevectors of the Dirac fermions $k_j$ are now discrete variables
with $k_j=2\pi n_j/(N+1)$ ($n_j=0,1,...,N$).
In terms of the conductivity this affects the scattering rate $\eta$. The latter
becomes $\eta_N$ in the finite sample and is determined by the equation
%\[
%g^{-1}=\frac{1}{(N+1)^2}
%\]
\beq
\sum_{n_1,n_2=0}^N\frac{1}{(N+1)^2(\eta_N^2+{\bar m}^2/4)+4\pi^2(n_1^2+n_2^2)}=\frac{1}{g}
\ .
\label{spe2}
\eeq
For a fixed parameter $\eta_N^2+{\bar m}^2/4$ the sum decreases with increasing $N$. 
This means that for given average gap ${\bar m}$ and variance $g$ the scattering rate $\eta_N$
decreases with increasing $N$. Consequently, the DC conductivity $\sigma_0$ of Eq. (\ref{cond5}) 
also decreases with increasing size $N$. If $\sigma_0>0$ for a given finite $N$ the conductivity can vanish
for a sufficiently large sample size. Whether or not it vanishes for the infinite sample depends on the 
asymptotic value $\eta$, which is determined by Eq. (\ref{spe1}). Thus the transition from
metallic to insulating behavior 
%is only defined for the infinite sample, whereas the behavior of the finite sample describes a crossover that 
depends on $N$, where the ability to conduct decreases with increasing $N$.
A special case is ${\bar m}=0$, where we have $\sigma_0=e^2/\pi h$ for any $N$, 
since $\eta_N$ does not enter the expression of $\sigma_0$ in Eq. (\ref{cond5}). 
This is in agreement with the results of the numerical calculation of 
Ref. \cite{beenakker10}.  
 
\begin{figure}[t]
\psfrag{mc}{$m_c$}
\psfrag{N}{$N$}
\begin{center}
\includegraphics[width=8cm,height=6cm]{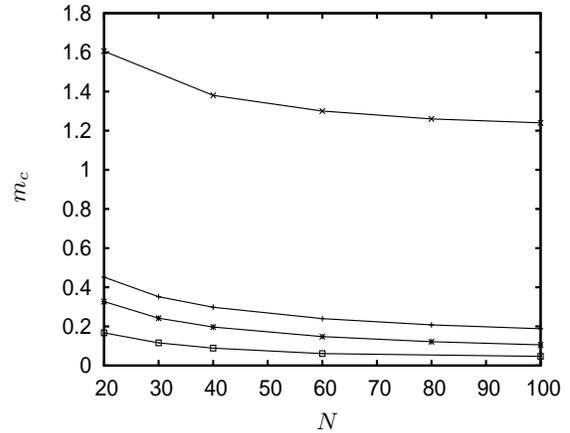}
\caption{Critical gap parameter $m_c=\sqrt{4\eta_N^2+{\bar m}^2}$ in units of the electronic hopping rate
of finite $N\times N$ samples as a function of $N$ for average gap
${\bar m}=0.06$ and for $g=2,4,5,10$. For fixed $N$ the critical gap parameter $m_c$ increases with 
increasing disorder strength $g$.
%in the presence of a gap since the scattering rate vanishes.
}
\label{finsp}
\end{center}
\end{figure}

\begin{figure}[t]
\psfrag{conductivity}{conductivity $(e^2/\pi h)$}
\psfrag{N}{$N$}
\begin{center}
\includegraphics[width=8cm,height=6cm]{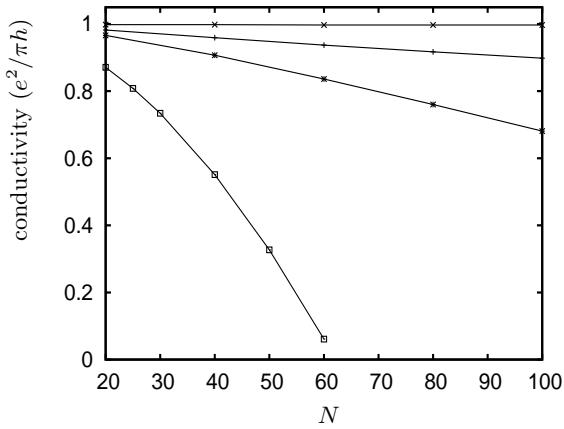}
\caption{Minimal conductivity $\sigma_0$ of finite $N\times N$ samples as a function of $N$ for average gap
${\bar m}=0.06$ and $g=2,4,5,10$. For fixed $N$ the conductivity increases with 
increasing disorder strength $g$.
%The asymptotic values of the infinite sample are
%$\sigma_0=0.98 e^2/\pi h$ ($g=10$), $\sigma_0=0.78 e^2/\pi h$ ($g=5$), and $\sigma_0=0$ ($g=2,4$).
}
\label{fincond}
\end{center}
\end{figure}

The behavior of the critical gap parameter $m_c$ and 
the conductivity $\sigma_0$ is presented
for sample sizes $N=20, ..., 100$ in Fig. \ref{finsp} and Fig. \ref{fincond}, respectively.
These results can be compared with the corresponding results for the infinite sample given in
Table \ref{table}. All calculations use the same average gap ${\bar m}=0.06$ for easy comparison. 

There are three remarkable facts for ${\bar m}>0$: 
(i) The conductivity and the scattering rate always decay monotonically with size $N$.
(ii) The slope of the nonzero conductivity decreases with increasing variance $g$.
In particular, the conductivity is almost constant for $g=10$ (cf. Fig. \ref{fincond}). 
(iii) At fixed $N$ the conductivity and the scattering rate increase always with increasing disorder $g$.
These results for the conductivity also agree with the findings of the recent numerical simulation of finite 
samples with random gap \cite{beenakker10}, although a quantitative comparison is not possible due to different
types of gap randomness: Instead of the single parameter $g$ of uncorrelated randomness in Eq. (\ref{spe2}),
the numerical simulation
uses correlated randomness with correlator $g_{r-r'}$. The identification of $g$ with the sum 
$K_0=\sum_r g_{r}$ may overestimate the strength of randomness in case of correlated disorder
\cite{majorana}. 
Moreover, our analytic result allows us to extrapolate to 
the infinite sample. This reveals an insulating (gapped) behavior for ${\bar m}>m_c$ but also a metallic 
behavior for ${\bar m}<m_c$, where $m_c$ is determined for the infinite sample (cf. Table \ref{table}).

The size dependence of the conductivity is similar if we send the width of the sample
to infinity and study a variable sample length $L$. Then one of the sums in Eq. (\ref{spe2}) 
becomes an integral that gives
 \[
\frac{1}{L+1}\sum_{n=0}^L \frac{\arctan(2\pi/\sqrt{{\bar \eta}^2+4\pi^2n^2/(L+1)^2})}
{2\pi\sqrt{{\bar \eta}^2+4\pi^2n^2/(L+1)^2}}=\frac{1}{g}
\ .
\]
The $g$ and $L$ (or $N$) dependence for the $N\times N$ square sample
and the infinite strip is plotted in Fig. \ref{fingraph1}.

\begin{table}
\begin{center}
\begin{tabular}{ccc}
\ \ \ \ \ $g$\ \ \ \ \  & $m_c$ & \ \ \ \ \ $\sigma_0$ ($e^2/\pi h$)  \ \ \ \ \ \\
2 & 0.06 & 0 \\
4 & 0.06 & 0 \\
5 & 0.09 & 0.78 \\
10 & 1.12 & 0.98 \\
\end{tabular}
\caption[smallcaption]{Critical gap parameter and minimal conductivity of the infinite sample
for average gap ${\bar m}=0.06$. (For this calculation the area of integration is approximated 
in Eq. (\ref{spe1}) by $k^2\le 1$.)}
\label{table}
\end{center}
\end{table}
 
\begin{figure}[t]
\psfrag{conductivity}{conductivity $(e^2/\pi h)$}
\psfrag{g}{$g$}
\psfrag{disorder}{$g$}
\begin{center}
\includegraphics[width=8cm,height=6cm]{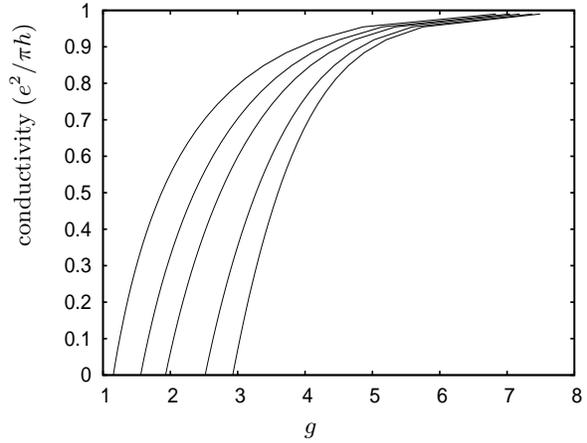}
\caption{Conductivity as a function of disorder strength $g$ for a fixed size $N\times N$ with $N=40,50,60,80,100$.
Larger values of $N$ require larger values of $g$ for the same conductivity.
}
\label{fingraph1}
\end{center}
\end{figure}
\begin{figure}[t]
\psfrag{conductivity}{conductivity $(e^2/\pi h)$}
\psfrag{g}{$g$}
\psfrag{disorder}{$g$}
\begin{center}
\includegraphics[width=8cm,height=6cm]{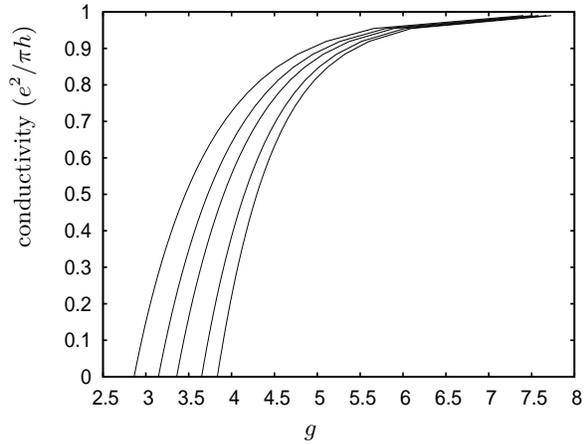}
\caption{Conductivity as a function of disorder strength $g$ for infinite width and finite lengths 
$L=40,50,60,80,100$. 
Larger values of $L$ require larger values of $g$ for the same conductivity.
}
\label{fingraph1}
\end{center}
\end{figure} 

Our observations in (i) -- (iii) provide a qualitative interpretation of the transport
properties in finite graphene samples in the presence of a random gap. First of all, the expression 
of $\sigma_0$ in Eq. (\ref{drude1}) implies that the conductivity depends on the random gap only 
through the average gap ${\bar m}$ and the scattering rate $\eta$. $\eta$ itself is determined
by Eq. (\ref{spe1}) (infinite sample) or by Eq. (\ref{spe2}) (finite sample), respectively,
as a function of ${\bar m}$ and of the variance of the gap fluctuations $g$.
There is no effective gap as long as the scattering rate is positive, since the conductivity is
nonzero. The fact that at 
fixed ${\bar m}$ the scattering rate increases with increasing $g$ is indicative of an increasing
density of states near the Dirac point due to disorder, since gap fluctuations allow the
appearence of additional states. It is less obvious whether these states are localized or
extended, although the increasing conductivity with increasing $g$ supports the latter.
The constant conductivity for ${\bar m}=0$ can be understood as a balanced interplay of an
increasing density of states near the Dirac point and a decreasing diffusion coefficient
due to gap fluctuations. Our scaling relation in Eq. (\ref{cond5}) yields a linear decay
of the conductivity when $m_c$ is approached. This is in agreement with a recent numerical simulation
for Dirac fermions with a uncorrelated random mass \cite{medvedyeva10}. It remains an open
question whether or not the linear behavior of the conductivity near $m_c$ is universal
with respect to different types of disorder or other details of the model on short scales. 
Numerical simulations for a similiar model have revealed that the conductivity obeys a 
power law near $m_c$ with a different exponent $\nu\approx 1.4$ \cite{kagalovksy10}.

In conclusion, we have found that transport in finite graphene samples with random gap
is enhanced by the gap fluctuations. In particular, the DC conductivity at fixed average gap
increases with increasing variance. The samples are metallic for sufficiently small but nonzero 
average gap. On the other hand, the conductivity always decreases with increasing sample size.

\begin{acknowledgments}
This project was supported by a grant from the Deutsche Forschungsgemeinschaft.
\end{acknowledgments}


\begin{thebibliography}{99}
\bibitem{saddlepoint}
This reflects the fact that $\eta$ is
a saddle point and Eq. (\ref{spe1}) is the corresponding saddle-point equation
\cite{ziegler98,ziegler09}.
\bibitem{majorana}
A corresponding simulation with an uncorrelated random Dirac mass has confirmed this
\cite{medvedyeva10}.
\end{thebibliography}
\end{document}